\documentclass[letterpaper]{article} % DO NOT CHANGE THIS
\usepackage{aaai2026}  % DO NOT CHANGE THIS
\usepackage{times}  % DO NOT CHANGE THIS
\usepackage{helvet}  % DO NOT CHANGE THIS
\usepackage{courier}  % DO NOT CHANGE THIS
\usepackage[hyphens]{url}  % DO NOT CHANGE THIS
\usepackage{graphicx} % DO NOT CHANGE THIS
\usepackage{natbib}  % DO NOT CHANGE THIS AND DO NOT ADD ANY OPTIONS TO IT
\usepackage{caption} % DO NOT CHANGE THIS AND DO NOT ADD ANY OPTIONS TO IT
\usepackage{algorithm}
\usepackage{algorithmic}
\usepackage{amssymb}
\usepackage{amsmath}
\usepackage{multirow}
\usepackage{booktabs}
\usepackage{xcolor}
\usepackage{newfloat}
\usepackage{listings}
\DeclareCaptionStyle{ruled}{labelfont=normalfont,labelsep=colon,strut=off} % DO NOT CHANGE THIS
\floatstyle{ruled}
\newfloat{listing}{tb}{lst}{}
\floatname{listing}{Listing}
\title{AV-Edit: Multimodal Generative Sound Effect Editing via \\ Audio-Visual Semantic Joint Control}
\author{
    Xinyue Guo\textsuperscript{\rm 1}, Xiaoran Yang\textsuperscript{\rm 2}, Lipan Zhang\textsuperscript{\rm 1}, Jianxuan Yang\textsuperscript{\rm 1}\textsuperscript{\rm}\thanks{corresponding author}, Zhao Wang\textsuperscript{\rm 3}, Jian Luan\textsuperscript{\rm 1}\\
}
\affiliations{
    \textsuperscript{\rm 1}MiLM Plus, Xiaomi Inc., China\\
    \textsuperscript{\rm 2}School of Electronic Information, Wuhan University, Wuhan, China\\
    \textsuperscript{\rm 3}Mi Ecosystem, Xiaomi Inc., China\\
    guoxinyue1@xiaomi.com, yxr888@whu.edu.cn, \{zhanglipan,yangjianxuan,wangzhao3,luanjian\}@xiaomi.com
}

\begin{document}

\maketitle

\begin{abstract}

Sound effect editing—modifying audio by adding, removing, or replacing elements—remains constrained by existing approaches that rely solely on low-level signal processing or coarse text prompts, often resulting in limited flexibility and suboptimal audio quality. To address this, we propose AV-Edit, a generative sound effect editing framework that enables fine-grained editing of existing audio tracks in videos by jointly leveraging visual, audio, and text semantics. Specifically, the proposed method employs a specially designed contrastive audio-visual masking autoencoder (CAV-MAE-Edit) for multimodal pre-training, learning aligned cross-modal representations. These representations are then used to train an editorial Multimodal Diffusion Transformer (MM-DiT) capable of removing visually irrelevant sounds and generating missing audio elements consistent with video content through a correlation-based feature gating training strategy. Furthermore, we construct a dedicated video-based sound editing dataset as an evaluation benchmark. Experiments demonstrate that the proposed AV-Edit generates high-quality audio with precise modifications based on visual content, achieving state-of-the-art performance in the field of sound effect editing and exhibiting strong competitiveness in the domain of audio generation.

\end{abstract}

\begin{links}
    \link{Extended version and code}{https://github.com/V2AResearch/AV-Edit}
\end{links}

\section{Introduction}

Recent advances in diffusion-based generative models have driven substantial progress in video editing, enabling diverse and controllable modifications while preserving the creative potential of generative synthesis~\cite{Feng_2024_CVPR,yang2025videograin}. Meanwhile, video-to-audio (V2A) generation~\cite{NEURIPS2023_98c50f47,xu2024video,gramaccioni2024stable} has attracted increasing attention, with research focusing on temporal synchronization and semantic alignment between generated audio and video. Despite these developments, sound effect editing—the task of modifying existing audio by adding, removing, or replacing elements—remains significantly less mature than video editing, particularly when precise synchronization with visual cues is required. Real-world cases—such as background noise unrelated to the scene contaminating the soundtrack of a live-action video, or objects being added or removed during visual editing without corresponding audio adjustments—underscore the need for seamless audio editing to ensure coherent multimedia experiences.

\begin{figure}[t]
	\centering
	\includegraphics[width=0.8\columnwidth]{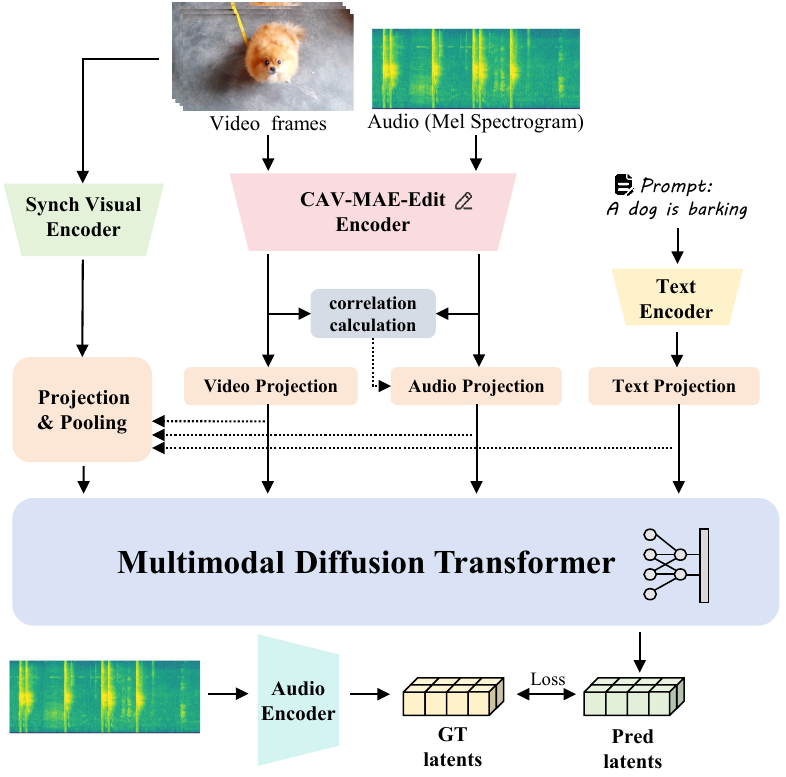}
	\caption{\textbf{Overview of the AV-Edit framework.} The pre-trained CAV-MAE-Edit encoder extracts joint audio–visual features, which are then fed into a multimodal diffusion model to generate the edited audio. }
	\label{fig1}
\end{figure}

Traditional sound effect editing methods rooted in signal processing manipulate waveforms, frequency spectra, or time–frequency features using operations such as cropping, splicing, and adaptive filtering. These techniques, while effective for basic modifications, rely heavily on manual parameter tuning and offer limited semantic control. Recent deep learning approaches attempt to overcome these limitations by employing diffusion models to edit audio conditioned on textual prompts~\cite{wang2023audit,manor2024zeroshot}, where the audio is encoded into a latent space and regenerated under prompt guidance. However, these methods rely on manual adjustments and prior knowledge of signal characteristics, struggling to dynamically synchronize audio edits with visual content while preserving original audio fidelity and audio–visual alignment.

To address these challenges, we introduce AV-Edit, a multimodal generative framework for fine-grained and visually aligned sound effect editing task. AV-Edit is designed to process audio, visual, and textual inputs, leveraging these multimodal cues within a generative model to achieve precise semantic alignment and temporal synchronization between the edited audio and the accompanying video. As illustrated in Figure~\ref{fig1}, the framework consists of four primary components: a joint audio–visual encoder (CAV-MAE-Edit), a text encoder~\cite{radford2021learning}, a visual synchronization encoder~\cite{iashin2024synchformer}, and a multimodal diffusion transformer (MM-DiT)~\cite{peebles2023scalable}. Among these, CAV-MAE-Edit is trained using a combination of contrastive pre-training and masked self-encoding. It learns joint audio–visual representations by projecting both modalities into a shared feature space, and employs mixing-based coding method with cross-attention to visual semantics applied during audio encoding/decoding, producing audio embeddings that are explicitly aligned with the visual content. The MM-DiT then leverages these embeddings, together with an audio–visual correlation-based feature gating training strategy, to guide the audio generation process and perform the desired editing operations. 

For evaluation, we construct VGG-Edit, a benchmark dataset derived from the VGGSound~\cite{chen2020vggsound} test set. We select 450 clips and systematically apply three types of edits—sound effect addition, deletion, and replacement—to create standardized evaluation scenarios for video-based sound effect editing. Both objective and subjective evaluations demonstrate that AV-Edit achieves state-of-the-art performance, producing high-quality, visually consistent audio while remaining competitive across multiple metrics in both sound effect editing and audio generation fields.

In summary, our main contributions are as follows:
\begin{itemize}
	\item A new generative paradigm for sound effect editing. AV-Edit jointly leverages audio, visual, and textual semantics to enable fine-grained, visually synchronized sound effect editing in videos.
	\item A contrastive audio–visual encoder tailored for sound effect editing. CAV-MAE-Edit integrates contrastive pre-training and masked self-encoding, to learn semantically relevant joint audio-visual representations.
	\item A benchmark for video-driven sound effect editing to facilitate fair comparison. VGG-Edit is a high-quality benchmark dataset featuring systematically constructed addition, deletion, and replacement scenarios.
\end{itemize}

\begin{figure*}[t]
	\centering
	\includegraphics[width=1.0\textwidth]{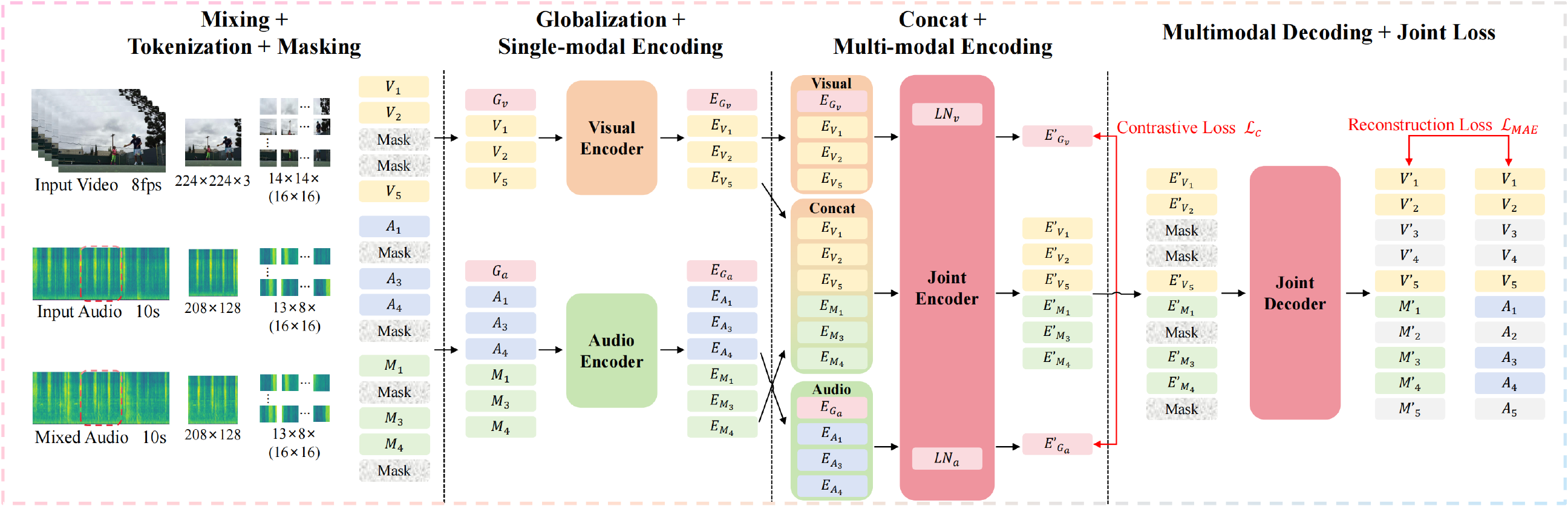} % Reduce the figure size so that it is slightly narrower than the column.
	\caption{\textbf{Overview of CAV-MAE-Edit network.} The single-modal encoders encode the visual and audio inputs separately. The multi-modal encoder and decoder process the joint embeddings of vision and audio.}
	\label{fig2}
\end{figure*}

\section{Related Work}

\subsection{Audio Editing}

Research on audio editing spans both traditional signal processing and modern generative methods. Early techniques, such as waveform editing tools~\cite{derry2012pc} and the WSOLA algorithm~\cite{verhelst1993overlap}, enable operations like cutting, splicing, time-stretching, and pitch-shifting. While suitable for basic modifications, these approaches rely heavily on manual parameter tuning and cannot synthesize content beyond the original recording.
With the advent of deep learning, generative models have introduced a new paradigm for audio editing. Instruction-driven systems~\cite{wang2023audit,vyas2023audiobox,han2023instructme} train task-specific generative models to add, replace, or mix audio segments, but require extensive annotated datasets. 
Other works~\cite{liu2023audioldm,manor2024zeroshot} employ Denoising Diffusion Probabilistic Models (DDPM) and Denoising Diffusion Implicit Models (DDIM) to guide edits from latent noise using text prompts; however, their reliance on precise textual descriptions limits applicability to more flexible or context-driven scenarios. 
Beyond text-conditioned methods, localized spectrogram-based editing has also been explored. For example, AudioMorphix~\cite{liang2025audiomorphix} supports spectrogram modifications but requires target and reference audio, while linguistically guided audio-visual editing~\cite{Liang_2024_ACCV} depends on handcrafted feature mappings, constraining scalability. 
Overall, current approaches struggle to achieve fine-grained, context-aware edits that dynamically adapt to video.

\subsection{Audio-Visual Pre-training}

Video-based sound effect editing requires robust joint representations of audio and video to ensure semantic alignment and temporal synchronization. Foundational works like CLIP~\cite{radford2021learning} and CLAP~\cite{laionclap2023} align text with images and audio, respectively, laying the groundwork for cross-modal alignment. Beyond text-based alignment, several studies focus on video–audio representation learning. For example, Synchformer~\cite{iashin2024synchformer} and DiffAV~\cite{luo2023diff} adopt contrastive learning to obtain temporally and semantically aligned audio–visual features.
Other approaches exploit feature fusion to enhance cross-modal understanding. UAVM~\cite{gong2022uavm} and VALOR~\cite{liu2024valor} integrate complementary information by extracting features from each modality and sharing parameters across specific network layers, thereby promoting modality interaction.
In parallel, masked signal modeling serves as a representation learning approach, where a large portion of the input data—such as images or audio—is masked out. This compels the encoder to learn rich semantic representations of the underlying signal in order to reconstruct the masked modality. 
Recent works~\cite{gong2022contrastive,georgescu2023audiovisual,araujo2025cav} combine masked autoencoding with contrastive learning to jointly capture modality-specific features and cross-modal relationships, enabling the development of robust audio–visual models suited for generative editing tasks.

\subsection{Multimodal Audio Generation}

Text-driven approaches~\cite{hung2024tangofluxsuperfastfaithful,lee2024etta} synthesize audio from textual descriptions, while video-to-audio approaches~\cite{zhang2024foleycrafter,viertola2025temporally} generate sound effects directly from visual inputs.
Some methods, such as SSV2A~\cite{SSV2A}, locally perceive multimodal sound sources from a scene with visual detection and cross-modality translation, address V2A generation at the sound-source level. Others pursue fully joint modeling across text, video, and audio: MMAudio~\cite{cheng2025mmaudio} and MultiFoley~\cite{chen2025video} integrates all three modalities into Transformer, enable flexible control of audio generation across modalities.
From an architectural perspective, autoregressive strategies~\cite{viertola2025temporally}, latent diffusion models (LDMs) and diffusion transformers (DiTs)\cite{luo2023diff,cheng2025mmaudio}, and rectified flow matching\cite{wang2024frieren} have all been explored.
While these methods advance multimodal generation, most focus on generating audio from silent videos. Editing sound effects in videos that already contain original audio—requiring both fine-grained control and preservation of temporal and semantic consistency—remains a largely unsolved challenge.

\section{Methodology}

The implementation of AV-Edit consists of two core stages: the pre-training stage of the audio-visual co-encoder CAV-MAE-Edit, and the training stage of the generative sound effect editing model implemented through modal-correlation feature gating and the MMDiT architecture. The overall structure of this framework is shown in Figure~\ref{fig1}.

\subsection{Editorial Audio-Visual Co-Encoder}

To facilitate signal reconstruction through joint semantic learning of visual and auditory inputs while acquiring unified audio-visual representations, we adopt the Contrastive Audio-Visual Masked Autoencoder (CAV-MAE) ~\cite{gong2022contrastive} framework for audio-visual pre-training. Furthermore, departing from conventional approaches, we innovatively introduce spectrogram segmenting and audio mixing strategies: the former can enhance the granularity of audio-visual pairs, while the latter enables the model to extract visually relevant information from the mixed audio based on visual semantics and encode the required information in an editing manner. Figure~\ref{fig2} illustrates the overall framework of our proposed method.

We shall commence with a review of the fundamental principles of CAV-MAE. For a given audio-visual pair, CAV-MAE initially crops the visual input to a fixed size and extracts the Mel spectrogram of the corresponding audio signal. Subsequently, the multi-modal signals denoted as \(\{v_{i} ,a_{i}\}\) undergo tokenization to generate corresponding token sequences \(\{V_{i} ,A_{i}\}\), which are then projected into the \( \mathbb{R}^{768} \) space via the modality-specific linear projection layers. Meanwhile, the model incorporates 2-D positional embeddings \((p_{v} ,p_{a})\) and modality-type embeddings \((m_{v} ,m_{a})\) into the aforementioned patch tokens. Ultimately, 75\% of the content for each modality is masked. The input to the encoder is formulated as follows:
\begin{equation}\label{eq1}
\begin{aligned}
&V^{\mathrm{unmask}}_{i}=\mathrm{Mask_{0.75}}(\mathrm{Proj_{v}}(\mathrm{Patch_{v}}(v_{i}))+p_{v}+m_{v}), \\
&A^{\mathrm{unmask}}_{i}=\mathrm{Mask_{0.75}}(\mathrm{Proj_{a}}(\mathrm{Patch_{a}}(a_{i}))+p_{a}+m_{a}).
\end{aligned}
\end{equation}
The sequences of visual and audio tokens are then forwarded through single-modal encoders \(E_{v}(\cdot )\) and \(E_{a}(\cdot )\), respectively, for learning modality-specific representations \(E_{V_{i}}\) and \(E_{A_{i}}\). All single-modal encoders employ the Vision Transformer (ViT) ~\cite{dosovitskiy2021an} architecture without sharing weights. Subsequently, \(E_{V_{i}}\), \(E_{A_{i}}\), and \(Concat(E_{V_{i}},E_{A_{i}})\) are fed into a joint audio-visual encoder, yielding \(E_{V^{'}_{i}}\), \(E_{A^{'}_{i}}\), and \(Concat(E_{V^{''}_{i}},E_{A^{''}_{i}})\). 
These three streams incorporate modality-specific layer normalization while sharing weights of other network layers in this multimodal encoder. The model leverages the single-modal outputs \(E_{V^{'}_{i}}\) and \(E_{A^{'}_{i}}\) for contrastive learning, and adopts the multimodal output \(Concat(E_{V^{''}_{i}},E_{A^{''}_{i}})\) for the reconstruction task. The contrastive loss is defined as:
\begin{equation}\label{eq2}
\mathcal{L}_{c}=-\frac{1}{N}\sum_{i=1}^{N}\log{(\frac{exp(s_{i,i}/\tau )}{ {\textstyle \sum_{k\ne i}exp(s_{i,k}/\tau)} +exp(s_{i,i}/\tau)})},
\end{equation}
where \(s_{i,j}=\left \| E_{V^{'}_{i}} \right \| ^{T}\left \| E_{A^{'}_{i}} \right \|\), and \(\tau \) is the temperature parameter of the similarity distribution. On the other hand, the reconstruction loss is defined as the mean squared error between the original patches \((V_{i}, A_{i})\) and the reconstructed patches \((V^{'}_{i}, A^{'}_{i})\), where the latter are the outputs of \(Concat(E_{V^{''}_{i}}, E_{A^{''}_{i}})\) after the multimodal decoder:
\begin{equation}\label{eq3}
\begin{aligned}
\mathcal{L}_{mae}=&\frac{1}{N}\sum_{i=1}^{N}(\frac{\sum (V'^{\mathrm{mask}}_{i}-\mathrm{norm}(V^{\mathrm{mask}}_{i}) )^{2}}{\left | V^{\mathrm{mask}}_{i} \right | } \\
&+\frac{\sum (A'^{\mathrm{mask}}_{i}-\mathrm{norm}(A^{\mathrm{mask}}_{i}) )^{2}}{\left | A^{\mathrm{mask}}_{i} \right | } ) ,
\end{aligned}
\end{equation}
where \(N\) denotes the batch size, and \(\left | V^{\mathrm{mask}}_{i} \right |\) as well as \(\left | A^{\mathrm{mask}}_{i} \right |\) represent the number of masked visual and audio patches, respectively. Finally, CAV-MAE weights and aggregates the contrastive loss and reconstruction loss to yield the final loss function:
\begin{equation}\label{eq4}
\mathcal{L}_{\mathrm{CAV-MAE}}=\lambda_{c}\mathcal{L}_{c}+\lambda_{mae}\mathcal{L}_{mae}.
\end{equation}
With respect to our proposed model, to acquire the target audio-visual co-encoder, we introduce the following improvements within the framework of CAV-MAE.
\begin{itemize}
	\item \textbf{Segment the spectrogram to increase the fine-grained nature of audio-visual pairs.} CAV-MAE maps the entire audio clip to a random frame of the video, failing to achieve accurate alignment between audio-visual events. Consequently, we preform frame-wise segmentation on audio-visual pairs, realizing manual temporally alignment on one hand, and ensure the semantic relevance of the two by narrowing the time range on the other.
    Specifically, frames are extracted from the video at a sampling rate of 8 FPS, where each video frame is paired with a \(0.125s\) audio segment cropped from the full Mel spectrogram at the corresponding temporal position. This approach enables fine-grained segmentation of the entire audio while preserving a certain level of semantic information, thereby achieving accurate alignment with the corresponding visual information.
    Let \(T\) denote the total number of video frames, and \(L\) represent the total length of the Mel spectrogram. The temporal center of the audio segment corresponding to the \(i\)-th frame is given by \(i*L/T\). Given that the duration of the audio segment for each frame is fixed as \(l\), the start and end positions of the audio segment corresponding to the \(i\)-th frame within the Mel spectrogram are defined as \(l_{start}=i*L/T-l/2\) and \(l_{end}=i*L/T+l/2\), respectively. This method enables the encoder to be trained with more precisely aligned audio-visual pairs.
	\item \textbf{Mixing to compel the model to learn audio-visual related semantics.} To encode features of audio segments that are relevant to visual semantics and achieve the desired editing effect, we mix the target audio with irrelevant audio, as detailed in Appendix B1.
    As shown in Figure~\ref{fig2}, we mask the same positions of the original and mixed audio, concatenate their outputs, and feed the result into an audio encoder. Subsequently, the mixed audio component within the audio embeddings is concatenated with the visual embeddings and input to the joint blocks of the multimodal encoder, where inter-modal information interaction is performed through Attention and MLP layers to derive semantically relevant audio-video embeddings. At the same time, the original audio embeddings, together with the duplicated visual embeddings, are encoded through the single-modal blocks within the multimodal encoder. This entire process can be represented as:
    \begin{equation}\label{eq5}
    \begin{aligned}
    &(e^{'}_{V},e^{'}_{M})=E_{j}(concat(e_{V},e_{M}));LN1_{av};LN2_{av}), \\
    &e^{''}_{V}=E_{v}(e_{V};LN1_{v};LN2_{v}), \\
    &e^{'}_{A}=E_{a}(e_{A};LN1_{a};LN2_{a}).
    \end{aligned}
    \end{equation}
    We input \((e^{'}_{V},e^{'}_{M})\) into the joint decoder to reconstruct data of each modality. For the reconstruction loss, we calculate it using the decoded audio-visual patches and the original audio-visual patches.   
	\item \textbf{Adding modal related global tokens to focus on global information.} Referring to~\cite{araujo2025cav}, we similarly introduce the global tokens \(G_{v}\) and \(G_{a}\), which continuously aggregate information during the single-modal encoding and multimodal encoding phases, focusing on the global representation of the respective modality. These encoded global tokens \(E^{'}_{G_{v}}\) and \(E^{'}_{G_{a}}\) are also ultimately used to calculate the contrastive loss.
\end{itemize}

\renewcommand{\arraystretch}{1}
\begin{table*}[htbp]
\centering
\begin{tabular}{ccccccc}
\hline
\multicolumn{3}{c}{\textbf{Method}} & \multicolumn{4}{c}{\textbf{Metrics}} \\
\cmidrule(lr){1-3}
\cmidrule(lr){4-7}
Segmenting & Global Token & Mixing & Audio MAE Loss & Visual MAE Loss & Constrastive Loss & Total Loss \\
\hline
\(\times\) & \(\times\) & \(\times\) & 2.219 & 1.335 & 1.484 & 3.569 \\
\(\checkmark\) & \(\times\) & \(\times\) & 0.450 & 1.097 & 8.552 & 1.632 \\
\(\checkmark\) & \(\checkmark\) & \(\times\) & \textbf{0.400} & 0.942 & \textbf{0.807} & 1.350 \\
\(\checkmark\) & \(\checkmark\) & \(\checkmark\) & 0.661 & \textbf{0.619} & 1.701 & \textbf{1.297} \\
\hline
\end{tabular}
\caption{Comparison of the losses for CAV-MAE-Edit under different methods.}
\label{table1}
\end{table*}
\renewcommand{\arraystretch}{1}

\subsection{Generative Sound Effect Editing}

With the aforementioned editorial audio-visual co-encoder in place, we next employ a multimodal diffusion model for generative sound effect editing. AV-Edit, as a generative sound effect editing model, can be divided into two parts: feature extraction and diffusion-based generation, which will be elaborated in detail below.

\subsubsection{Feature Extraction}

Audio and visual semantic embeddings \(F_{a}\) and \(F_{v}\) incorporating both temporal and spatial dimensions are extracted via the pre-trained co-encoder CAV-MAE-Edit. These embeddings are flattened along the temporal dimension and pooled across the spatial dimension. 
Subsequently, a correlation-based feature gating strategy is adopted to compute the cosine similarity between audio and visual embeddings frame-by-frame, thereby determining the degree of semantic alignment between the current image and audio features. This mechanism enables the determination of whether to feed the audio features of the current frame into the generative network. The proposed algorithm retains audio embeddings whose similarity scores exceed a predefined threshold \(r\), while substituting semantically irrelevant audio features with empty features. 
The textual feature \(F_{t}\) is extracted using CLIP, and the visual synchronization feature \(F_{sync}\) is obtained through Synchformer to facilitate the generation of video-synchronized audio. Thereafter, the visual-audio-textual semantic features are projected onto the predefined dimensions by their respective projection layers; on the one hand, they serve as inputs to the diffusion model, and on the other hand, they undergo pooling across time and are summed with the temporal embedding to form global control information \(g_{c}\). 
Furthermore, the visual synchronization features, after projection, are added to \(g_{c}\) to form the temporal synchronization information \(f_{c}\), which is incorporated into the generation process. During the training process of AV-Edit, all these encoders are frozen, and the projection layer will be trained from scratch.

\subsubsection{Diffusion Generation}

Aiming to fully leverage multimodal information, and thus, similar to MMAudio, we adopt the MM-DiT block from ~\cite{esser2024scaling}. The detailed structure of the generative model component in AV-Edit is provided in Appendix A1. 
Within the diffusion model, initially, the target audio is encoded into latent space as \(x\) by a pre-trained VAE encoder, followed by forward diffusion. Correspondingly, the VAE decoder leverages the backward denoising process to obtain the output \(\hat{x} \), namely the high-dimensional representation retrieved from the latent space.
During the forward diffusion, Gaussian noise is added to the latent \(x_{0}\) of the clean audio to obtain noisy latent \(x_{t}\) at different time steps \((t\in \{1,\dots ,T\})\). In step t, \(x_{t}=\sqrt{\alpha _{t}}\cdot x_{t-1}+ \sqrt{1-\alpha _{t}}\cdot \epsilon _{t}\), where \(\epsilon _{t}\sim \mathcal{N}(0,\mathrm{\mathbf{I} } )\) is Gaussian noise and \(\alpha _{t}\) is a scale parameter. Correspondingly, the inverse process is an iterative denoising of pure noise \(x_{T}\) to recover the original data \(x_{0}\). 
It is well-known that DiT adopts Transformer as the backbone network of the diffusion model to modeling the noise prediction function \(\epsilon _{\theta }\). Furthermore, the multimodal diffusion model we use incorporates control conditions into the noise prediction process. Specifically, it concatenates the queries, keys, and values of the visual-audio-textual embeddings with those of the latent \(x_t\), and then fuses features from these modalities through joint attention. 
The predicted latent can be expressed as \(\epsilon _{\theta}(x_{t},t,(f_{a},f_{v},f_{t}))\) and the network is trained with the goal:
\begin{equation}\label{eq6}
\mathcal{L}(\theta )=\mathbb{E}_{x_{0},\epsilon,t}[\left \|  \epsilon -\epsilon _{\theta}(x_{t},t,(f_{a},f_{v},f_{t}))\right \| ^{2}],
\end{equation}
where \(\epsilon \sim \mathcal{N}(0,\mathrm{\mathbf{I} })\) and \(t\sim \mathrm{Uniform(1,T)}\). Detailed implementation specifications for the remaining modules of the generative network are provided in Appendix A2. Furthermore, the Classifier-Free Guidance (CFG) technique is incorporated during the training process. Specifically, in the training phase, the visual embedding \(f_{v}\), audio embedding \(f_{a}\), and text embedding \(f_{t}\) are randomly masked with learnable empty embeddings \(\varnothing_{v}\), \(\varnothing_{a}\), and \(\varnothing_{t}\) respectively, with a masking probability of 10\%. The outputs of the model under conditional and unconditional settings is balanced by a guidance scale \(s\), and the adjusted predicted latent variable is formulated as follows:
\begin{equation}\label{eq7}
\begin{aligned}
\hat{\epsilon} _{\theta}(x_{t},t,(f_{a},f_{v},f_{t})) =&s\cdot \epsilon_{\theta}(x_{t},t,(f_{a},f_{v},f_{t}))\\
&+(1-s)\epsilon_{\theta}(x_{t},t,(\varnothing_{a},\varnothing_{v},\varnothing_{t})).
\end{aligned}
\end{equation}
Classifier-Free Guidance serves as an effective approach to enhancing generation performance and editing capabilities, while enabling a balanced trade-off between the quality and diversity of the final output.

\renewcommand{\arraystretch}{1}
\begin{table*}[htbp]
\centering
\setlength{\tabcolsep}{4.5pt} % 调整列距
\begin{tabular}{lccccccc}
\hline
\multirow{2}{*}{\textbf{Method}} & \multicolumn{2}{c}{\textbf{Distribution matching}} & \multicolumn{1}{c}{\textbf{Audio quality}} & \multicolumn{1}{c}{\textbf{Semantic align}} & \multicolumn{1}{c}{\textbf{Temporal align}} & \multicolumn{2}{c}{\textbf{Efficiency}} \\
\cmidrule(lr){2-3}
\cmidrule(lr){4-4}
\cmidrule(lr){5-5}
\cmidrule(lr){6-6}
\cmidrule(lr){7-8}
 & \(\mathrm{KL_{PANN_{s}}}\downarrow\) & \(\mathrm{KL_{PaSST}}\downarrow\) & IS\(\uparrow\) & IB\(\uparrow\) & DeSync\(\downarrow\) & Params & Time(s) \\
\hline
Seeing\&Hearing & 2.92 & 2.93 & 6.19 & \textbf{33.85} & 1.193 & 415M & 30.18 \\
FoleyCrafter & 2.30 & 2.25 & 15.67 & 25.75 & 1.222 & 1.22B & 3.71 \\
V-AURA\(^{\star}\) & 2.44 & 2.08 & 10.29 & 27.64 & 0.813 & 695M & 33.63 \\
Frieren\(^{\star}\) & 3.36 &  3.59 & 8.37 & 17.31 & 1.259 & 159M & 2.68 \\
V2A-Mapper\(^{\star\Diamond}\) & 2.69 & 2.55 & 12.47 & 22.55 & 1.225 & 229M & - \\
AudioX & 1.99 & 1.86 & 16.82 & 25.86 & 1.237 & 1.10B & 13.00 \\
MMAudio-S & \textbf{1.63} & 1.60 & 14.53 & 28.23 & 0.505 & 157M & 2.56 \\
MMAudio-L & \textbf{1.63} & 1.46 & 18.13 & \underline{32.73} & \underline{0.465} & 1.03B & 3.01 \\
\hline
AV-Edit-S (ours) & 1.71 & 1.61 & 20.38 & 28.08 & 0.480 & 178M & 2.69\\
AV-Edit-L (ours) & \underline{1.67} & \textbf{1.41} & \underline{22.48} & 31.68 & \textbf{0.441} & 1.15B & 3.18\\
AV-Edit w/o Audio & 1.73 & \underline{1.45} & \textbf{23.08} & 30.73 & 0.481 & 1.15B & 3.16\\
\hline
\end{tabular}
\caption{Audio Generation results on the VGGSound test set. The number of parameters excludes all pre-trained encoders. All samples were generated on an H800 GPU. \(\star\): does not use text input during testing. \(\Diamond\): evaluated using generation samples obtained directly from the authors. The best result for each indicator is bolded and the second best result is underlined.}
\label{table2}
\end{table*}
\renewcommand{\arraystretch}{1}

\section{Experiments}

In this section, we perform comprehensive experiments and present various evaluation results.

\subsection{Audio-Visual Co-Encoder CAV-MAE-Edit}

The proposed CAV-MAE-Edit undergoes pre-training to acquire joint audio-visual representations and the signal reconstruction ability corresponding to each modality.

\subsubsection{Dataset} 

We utilize the complete VGGSound dataset for training the proposed encoder. VGGSound contains 200K 10-second videos spanning 309 categories, with each video containing a semantically explicit audio-visual pair. For each video, we extract video frames at a sampling rate of 8 FPS and the complete audio at a sampling rate of 16 kHz. Eventually, 80 frames of video and a 128-dimensional Mel spectrogram with a length of 1024 are obtained.

\subsubsection{Implementation Details} 

We set the length of the Mel spectrogram segment corresponding to each video frame to 208, and slice the audio to obtain fine-grained visual-audio pairs. The video frame and audio clip are then segmented with a patch size of 16, resulting in 14 × 14 blocks of visual embeddings and 13 × 8 blocks of audio embeddings. Audio mixing is performed with a probability of 0.5, while audio and visual embeddings are masked with a probability of 0.5, resulting in three sets of data: mixed-masked audio, masked clean audio and masked visual embeddings. These three streams are fed into their respective encoders. Training is conducted on four NVIDIA H800 GPUs with a learning rate of 1e-4 and a batch size of 160 until convergence, which takes approximately 150 epochs. Furthermore, the weights parameters of the loss function in Equation 4 are set to \(\lambda_{c}=0.01\) and \(\lambda_{mae}=1\), respectively. The variation of various loss values during training is illustrated in Figure B of Appendix B2. 

\subsubsection{Ablation} 

To explore the effectiveness of the improvements proposed in this work for the audio-visual co-encoder with respect to CAV-MAE, we conducted ablation experiments under equivalent conditions. Table~\ref{table1} shows the impacts of spectrogram segmenting, adding global tokens, and audio mixing on the model. It can be observed that spectrogram segmentation significantly reduces the Audio MAE loss, while the incorporation of global tokens enhances the model’s capability in contrastive learning. Although the introduction of audio mixing increases the audio MSE loss, it enables the model to perform audio encoding in an edit-oriented manner while minimizing the overall model loss.

\subsection{Audio Generation Part of AV-Edit}

After completing the pre-training of the encoder, the audio is next generatively edited using the model AV-Edit.

\subsubsection{Experimental Setup}

In this part, we use the datasets VGGSound, Audiocaps~\cite{kim2019audiocaps}, Wavcaps~\cite{mei2024wavcaps}, and Clotho~\cite{drossos2020clotho} to train the model. Audio, visual, and textual features are extracted from these datasets, and for modalities lacking in the datasets, we set their features to learnable null tokens such as \(\varnothing_{a}\), \(\varnothing_{v}\) and \(\varnothing_{t}\). 
For the aforementioned audio-visual similarity threshold \(r\), we set it to 0.3. The detailed process of threshold selection is provided in Appendix A3.
The structure of the diffusion component, the same as in MMAudio, is divided into two types: small model and large model. The small model contains 4 MM-DiT blocks and 8 Single-Modal DiT blocks, generating 16 kHz audio encoded as 20-dimensional latent, while the large model contains 7 MM-DiT blocks and 14 Single-Modal DiT blocks, generating 44.1 kHz audio encoded as 40-dimensional latent. Using 25 inference steps and setting the classifier-free guidance strength to 4.5, the model is trained on four NVIDIA H800 GPUs for about 43 epochs.

\subsubsection{Evaluation Metrics} 

The generated audio is evaluated using distribution matching, audio quality, semantic alignment, and temporal alignment. Distribution matching is measured using the Kullback-Leibler (KL) distance of the features extracted by the PANNs~\cite{kong2020panns} and PaSST~\cite{koutini22passt} models. Audio quality is evaluated using the Inception Score, while the semantic alignment score IB and the temporal alignment score DeSync are calculated using ImageBind~\cite{girdhar2023imagebind} and Synchformer, respectively. Appendix C1 details how these evaluation indicators were calculated.

\begin{figure}[htbp]
	\centering
	\includegraphics[width=1.0\columnwidth]{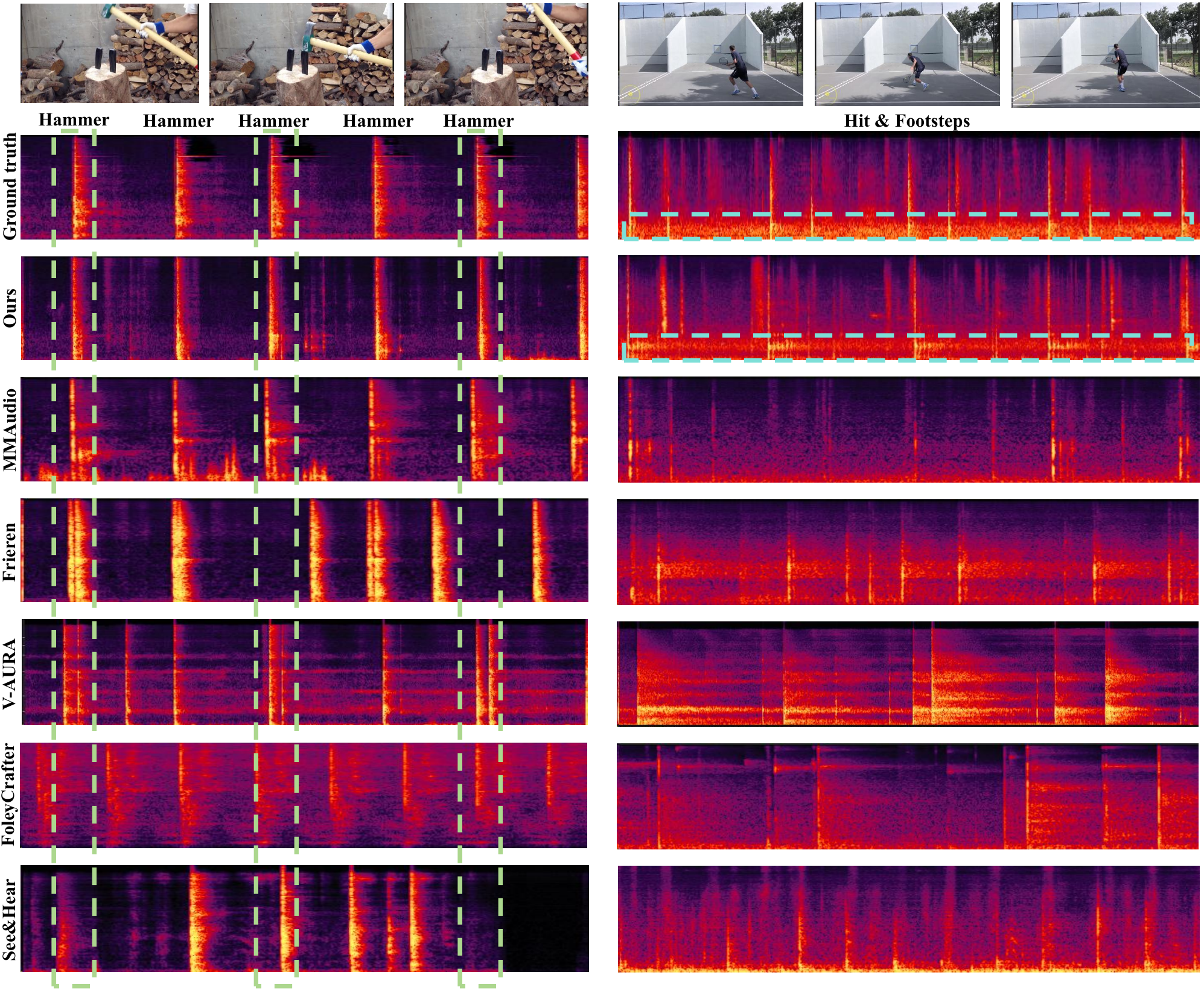}
	\caption{The spectrograms of generated audios.}
	\label{fig3}
\end{figure}

\subsubsection{Quantitative Results}

We evaluate the generative performance of the model using the test set of VGGSound, and the comparison models contained representative V2A models Seeing and Hearing~\cite{xing2024seeing}, FoleyCrafter~\cite{zhang2024foleycrafter}, V-AURA~\cite{viertola2025temporally}, Frieren~\cite{wang2024frieren}, V2A-Mapper~\cite{wang2024v2a}, AudioX~\cite{tian2025audiox}, and MMAudio. Unlike other models, the input of our model AV-Edit includes not only visual-textual features but also raw audio features, which, however, interferes with audio generation to some extent. The above models are used to generate audios with 8 seconds for testing, and the results are shown in Table~\ref{table2}. Our model is overwhelmingly superior in terms of generated audio quality, with top performance in distribution matching and temporal alignment, and semantic alignment second only to Seeing\&Hearing and MMAudio. It is worth noting that, as a generative sound effect editing model, our model has already outperformed existing state-of-the-art models in most key metrics and ranked the best when evaluated solely from the perspective of audio generation without the input of audio features. For a better comparison, the visualization results are shown in Figure~\ref{fig3} and analyzed in Appendix C2.

\subsection{Sound effect Editing Based on Video Content}

Lastly, we explore the editing performance of our proposed framework in detail, focusing on the addition, removal, and replacement of original audio tracks in videos.

\subsubsection{Dataset VGG-Edit}

To evaluate the model's capability of video-based sound effect editing, we design a dedicated dataset called VGG-Edit based on the test set of VGGSound. To be specific, we manually select 450 visually meaningful videos from approximately 15,000 test videos. For 150 of these videos, we add sound effects irrelevant to the visual semantics to their audio tracks; for another 150 videos, we remove the sound effects relevant to the visual semantics from their audio tracks; and for the remaining 150 videos, we replace the original visually semantic-relevant sound effects in their audio tracks with sound effects that are irrelevant.

\subsubsection{Objective Results}

We selecte the V2A models AudioX and DeepSound~\cite{liang2025deepsound} that support audio track input as comparative methods, and conducte tests on the proposed sound effect editing dataset, aiming to perform operations of adding, deleting, and replacing the audio tracks of the input videos. For AudioX, we input the videos with the original audio tracks and the specified textual descriptions. DeepSound handles this slightly differently: in the add and replace tasks, the videos with the original audio tracks and the specified textual descriptions are fed into step 1 of the model, while in the delete task, we skip the first step of the model and feed the aforementioned inputs into the second step of the model to separate the sound effects that are not related to the visual semantics. After extracting the features of the output audio and target audio using PANNs, we use the KL distance and Fréchet~\cite{frechet1906quelques} distance to evaluate the objective metrics of the editing performance of the aforementioned models, and the results are shown in Table~\ref{table3}. 

\renewcommand{\arraystretch}{1}
\begin{table}[htbp]
\centering
\setlength{\tabcolsep}{2.5pt} % 调整列距
\begin{tabular}{lcccccc}
\hline
\multirow{2}{*}{\textbf{Method}} & \multicolumn{2}{c}{\textbf{Addition}} & \multicolumn{2}{c}{\textbf{Removal}} & \multicolumn{2}{c}{\textbf{Replacement}}\\
\cmidrule(lr){2-3}
\cmidrule(lr){4-5}
\cmidrule(lr){6-7}
 & FD\(\downarrow\) & KL\(\downarrow\) & FD\(\downarrow\) & KL\(\downarrow\) & FD\(\downarrow\) & KL\(\downarrow\) \\
\hline
\textcolor{gray}{VGG-Edit} & \textcolor{gray}{131.63} & \textcolor{gray}{5.86} & \textcolor{gray}{121.07} & \textcolor{gray}{5.74} & \textcolor{gray}{135.96} & \textcolor{gray}{6.12} \\
\hline
AudioX w/ T & 47.76 & 2.60 & 51.71 & 2.86 & 79.64 & 3.66 \\
DeepSound w/ T & 40.69 & 2.53 & 75.16 & 3.63 & 126.10 & 5.91 \\
AV-Edit w/o T & 36.05 & 1.56 & 27.56 & 0.99 & 48.24 & 2.13 \\
AV-Edit w/ T & \textbf{31.04} & \textbf{1.39} & \textbf{25.24} & \textbf{0.95} & \textbf{38.00} & \textbf{1.54} \\
\hline
\end{tabular}
\caption{The objective results of various sound effect editing methods on the VGG-Edit dataset.}
\label{table3}
\end{table}
\renewcommand{\arraystretch}{1}
Our method outperforms others across all metrics, enabling purposeful sound effect addition, removal, and replacement to approximate the target audio. Compared to AudioX and DeepSound, it achieves significant improvements in removal and replacement tasks, with FD and KL distances more aligned with the target. Additionally, textual instructions enhance sound effect processing effectiveness.

\renewcommand{\arraystretch}{1}
\begin{table}[htbp]
\centering
\setlength{\tabcolsep}{5pt} % 调整列距
\renewcommand{\arraystretch}{1.1} 
\begin{tabular}{lccccc}
\hline
\multirow{1}{*}{\textbf{Method}} & \multicolumn{1}{c}{\textbf{F}\(\uparrow\)} & \multicolumn{1}{c}{\textbf{PQ}\(\uparrow\)} & \multicolumn{1}{c}{\textbf{C}\(\uparrow\)} & \multicolumn{1}{c}{\textbf{RS}\(\uparrow\)} & \multicolumn{1}{c}{\textbf{IA}\(\uparrow\)}\\
\hline
AudioX w/ T & 42.2 & 49.1 & 49.7 & 52.5 & 58.8 \\
DeepSound w/ T & 47.3 & 39.8 & 46.9 & 48.0 & 36.7 \\
AV-Edit w/o T & \textbf{64.4} & 58.8 & 63.2 & 63.3 & 68.0 \\
AV-Edit w/ T & 62.4 & \textbf{66.2} & \textbf{64.2} & \textbf{64.7} & \textbf{70.8} \\
\hline
\end{tabular}
\caption{The subjective results of various sound effect editing methods on the VGG-Edit dataset.}
\label{table4}
\end{table}
\renewcommand{\arraystretch}{1}
\subsubsection{Subjective Results}

In terms of subjective evaluation, we use the metrics proposed by the audio editing model AudioMorphix: fidelity (F), perceptual quality (PQ), consistency (C), regional specificity (RS), and instruction adherence (IA). The specific meanings and scoring methods of these metrics are given in Appendix D1. Table 4 demonstrates the superiority of our method AV-Edit, which can well refer to the features of the original video's audio track, perform sound effect addition or deletion operations based on visual semantics. The textual feature input has a guiding effect on the model, which improves the region specificity and instruction adherence, etc. of the edited audio with a slight reduction in fidelity. Figure~\ref{fig4} shows the visualization results of sound effect editing performed by AV-Edit and analyzed in Appendix D2. For example, it can be seen from the middle part of Figure~\ref{fig4} that the edited audio removes the sound effect associated with the eliminated image from the original track, and the retained portion is enhanced. At last, More ablation experiments can be seen in Appendix E.

\begin{figure}[htbp]
	\centering
	\includegraphics[width=1.0\columnwidth]{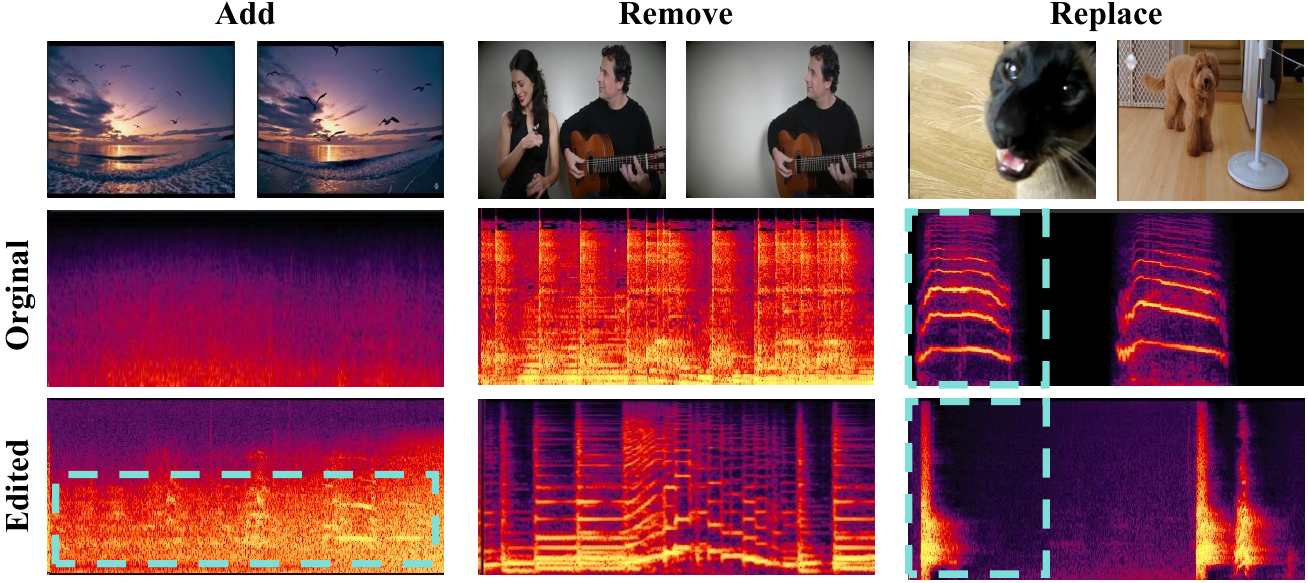}
	\caption{Examples of three editing manipulations: add, remove and replace.}
	\label{fig4}
\end{figure}

\section{Conclusion}

In this paper, we propose AV-Edit, a groundbreaking generative sound effect editing framework for fine-grained editing of existing audio tracks in videos. With a specially designed audio-visual co-encoder and a multimodal training strategy that introduces correlation-based feature gating, our approach achieves superior sound editing performance, removes visually irrelevant sounds, and generates missing sound effect elements. For evaluation, we construct VGG-Edit, a benchmark dataset for video-based sound effect editing. Through comprehensive experiments, we demonstrate that AV-Edit outperforms existing audio generation models and, simultaneously, fills the gap in video-based sound effect editing models. In addition, our approach also has a limitation - it does not preserve the original audio completely without distortion - and we hope to address this pain point in future work.

\section{Acknowledgements}

This work was supported by the MiLM Plus, Xiaomi Inc.

\bibliography{aaai2026}

\newpage

\section*{Appendix}

\subsection*{A. AV-Edit}

\subsubsection*{A1. Detailed structure of the generation section in AV-Edit}

Figure A shows the detailed structure of the generation section in AV-Edit, and its implementation details are described in section A2.

\subsubsection*{A2. Implementation details of the generation section in AV-Edit}

As shown in Figure A, the generation component of AV-Edit consists of N1 MMDiT blocks and N2 Single-modal DiT blocks. The former is used to fuse multimodal information, while the latter reconstructs the target solely based on audio latents. The structure of the Single-modal DiT block is the same as that of the MMDiT block, except that it removes the three streams \(F_{a}\), \(F_{t}\), and \(F_{v}\), converting the joint attention into self-attention. The model employs the conditional flow matching algorithm to model the generation process, defining a probabilistic path from the noise distribution to the data distribution, thereby enabling efficient generation and inference. The following is a top-down introduction to the implementation of each part of the model. To begin with, the encoders described in the main text are used to encode information from each modality, which is then projected to the specified dimension through their respective projection layers. Subsequently, global features \(g_{c}\) and \(f_{c}\) are injected into the network via adaptive normalization layers, and the audio and visual embeddings are aligned using RoPE embeddings to achieve temporal synchronization. The queries, keys, and values of embeddings from different modalities are concatenated, processed through joint attention, and the output is then sequentially partitioned into each modality. At last, operations such as ConvMLP are used to restore the intra-modal temporal structure.

\subsubsection*{A3. Selection of the feature gating threshold \(r\)}
We traverse all the cosine similarity of the audio-visual features in the VGGSound training set, and finally took its median value of 0.3 as the threshold \(r\).

\subsection*{B. CAV-MAE-Edit}

\subsubsection*{B1. Implementation details of audio mixing strategy in CAV-MAE-Edit}

We randomly select audio segments that are semantically irrelevant to the target audio and have matching lengths from the training set as distractor audio. First, we perform peak normalization on both the target audio and the distractor audio separately to eliminate interference caused by volume differences. Then, according to the preset signal-to-noise ratio (SNR) range (-5 dB to 15 dB), we calculate the scaling factor of the distractor audio and conduct linear mixing following the formula \(a_{m} = a_{t} + \alpha a_{d}\) (where \(a_{t}\) denotes the target audio, \(a_{d}\) represents the distractor audio, \(\alpha = \sqrt{10^{-\mathrm{SNR} /10} } \) and \(a_{m}\) is the mixed audio). Finally, we perform secondary peak normalization on the mixed audio to obtain the final mixed audio samples.

\subsubsection*{B2. Training process of CAV-MAE-Edit}
\begin{figure}[h]
    \renewcommand{\thefigure}{B}
	\centering
	\includegraphics[width=0.5\textwidth]{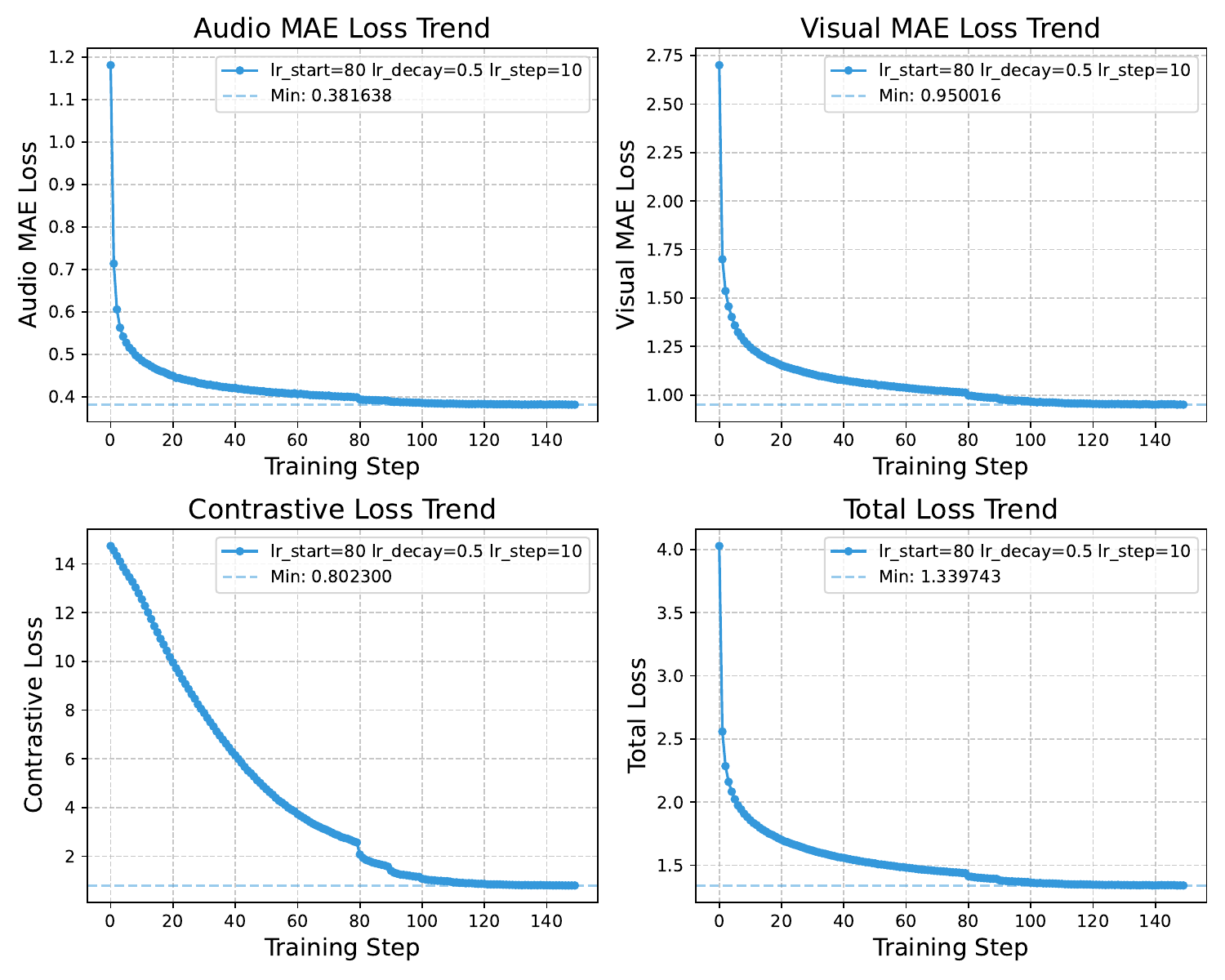}
	\caption{Curves of different loss functions during CAV-MAE-Edit training process}
	\label{figB}
\end{figure}

Figure B shows the variation curves of four loss during the training process of the encoder CAV-MAE-Edit, namely the audio reconstruction loss, visual reconstruction loss, audio-visual contrastive loss, and the weighted overall loss. We adopt a step decay strategy for the learning rate in training, with an initial learning rate of 1e-4. Starting from the 80th epoch, the learning rate is decayed to 50\% of its original value every 10 epochs. It can be observed that compared with the reconstruction loss, the contrastive loss decreases more slowly, and the overall training converges at the 150th epoch.

\subsection*{C. Evaluation of Audio Generation}

\subsubsection*{C1. Calculation of Evaluation Metrics}

\textbf{Distribution matching} refers to the distribution similarity between generated audio and target audio under a specific feature extraction model. We use the PANNs and PaSST models to extract audio features, and calculate the Kullback-Leibler distance between the generated audio and the target audio as the evaluation result for distribution matching. Among them, PANNs operates at 16 kHz, while PaSST operates at 32 kHz. \textbf{Audio quality} refers to the quality of the generated audio. We use the PANNs model to extract features from the generated audio, and then use the Inception Score to measure Audio quality. \textbf{Semantic alignment} refers to the semantic similarity between the generated audio and the input video. We use the ImageBind model to extract features of the generated audio and the input visual information, and calculate the cosine similarity between the two to obtain the semantic alignment score. \textbf{Temporal alignment} refers to the temporal synchronization between the generated audio and the input video. We use the Synchformer model to extract features of the generated audio and the input visual information, compare the temporal offsets between the two, and then obtain the temporal synchronization score DeSync.

\subsubsection*{C2. Visualization of Generated Audio}

Figure C shows the entirely visualization results of the audio generated by the model mentioned in the main text. It can be seen from the figure that the model we proposed has prominent advantages in terms of temporal synchronization, semantic diversity, and audio quality. For the left part of Figure C, the audio generated by our model is aligned with the real audio and free of extra noise. For the right part of Figure C, our model not only generates time-aligned ball-hitting sounds but also footstep sounds, demonstrating semantic diversity.

\renewcommand{\arraystretch}{1.15}
\begin{table*}[htbp]
\centering
\begin{tabular}{ccccccc}
\hline
Audio Encoder & Visual Encoder & \(\mathrm{KL_{PANN_{s}}}\downarrow\) & IS\(\uparrow\) & IB\(\uparrow\) & DeSync\(\downarrow\) \\
\hline
CLAP & CLIP & 1.73 & 14.09 & 28.13 & 0.563 \\
ImageBind & ImageBind & \textbf{1.05} & 15.28 & 29.24 & 0.572 \\
CAV-MAE-Edit & CAV-MAE-Edit & 1.67 & \textbf{22.48} & \textbf{31.68} & \textbf{0.441} \\
\hline
\end{tabular}
\caption{Quantitative results for the audio generation after replacing the encoder.}
\label{table1}
\end{table*}
\renewcommand{\arraystretch}{1}

\renewcommand{\arraystretch}{1.15}
\begin{table*}[htbp]
\centering
\begin{tabular}{cccccccc}
\hline
\multicolumn{2}{c}{\textbf{Method}} & \multicolumn{2}{c}{\textbf{Addition}} & \multicolumn{2}{c}{\textbf{Removal}} & \multicolumn{2}{c}{\textbf{Replacement}}\\
\cmidrule(lr){1-2}
\cmidrule(lr){3-4}
\cmidrule(lr){5-6}
\cmidrule(lr){7-8}
 Audio Encoder & Visual Encoder & FD\(\downarrow\) & KL\(\downarrow\) & FD\(\downarrow\) & KL\(\downarrow\) & FD\(\downarrow\) & KL\(\downarrow\) \\
\hline
CLAP & CLIP & 52.10 & 2.91 & 43.03 & 2.56 & 59.58 & 3.11 \\
ImageBind & ImageBind & 37.18 & 1.63 & 33.81 & 1.84 & 54.20 & 2.64 \\
CAV-MAE-Edit & CAV-MAE-Edit & \textbf{31.04} & \textbf{1.39} & \textbf{25.24} & \textbf{0.95} & \textbf{38.00} & \textbf{1.54} \\
\hline
\end{tabular}
\caption{Objective results for the sound effect editing after replacing the encoder.}
\label{table2}
\end{table*}
\renewcommand{\arraystretch}{1}

\renewcommand{\arraystretch}{1.15}
\begin{table*}[htbp]
\centering
\begin{tabular}{cccccc}
\hline
Correlation Calculation & \(\mathrm{KL_{PANN_{s}}}\downarrow\) & IS\(\uparrow\) & IB\(\uparrow\) & DeSync\(\downarrow\) \\
\hline
\(\times\) & 1.72 & 14.06 & 27.16 & 0.573 \\
\(\checkmark\) & \textbf{1.67} & \textbf{22.48} & \textbf{31.68} & \textbf{0.441} \\
\hline
\end{tabular}
\caption{Quantitative results for the audio generation after deleting correlation calculation.}
\label{table3}
\end{table*}
\renewcommand{\arraystretch}{1}

\renewcommand{\arraystretch}{1.15}
\begin{table*}[htbp]
\centering
\begin{tabular}{ccccccc}
\hline
\multirow{2}{*}{\textbf{Correlation Calculation}} & \multicolumn{2}{c}{\textbf{Addition}} & \multicolumn{2}{c}{\textbf{Removal}} & \multicolumn{2}{c}{\textbf{Replacement}}\\
\cmidrule(lr){2-3}
\cmidrule(lr){4-5}
\cmidrule(lr){6-7}
 & FD\(\downarrow\) & KL\(\downarrow\) & FD\(\downarrow\) & KL\(\downarrow\) & FD\(\downarrow\) & KL\(\downarrow\) \\
\hline
\(\times\) & 57.37 & 3.23 & 59.14 & 3.23 & 62.21 & 3.67 \\
\(\checkmark\) & \textbf{31.04} & \textbf{1.39} & \textbf{25.24} & \textbf{0.95} & \textbf{38.00} & \textbf{1.54} \\
\hline
\end{tabular}
\caption{Objective results for the sound effect editing after deleting correlation calculation.}
\label{table4}
\end{table*}
\renewcommand{\arraystretch}{1}

\subsection*{D. Evaluation of Sound Effect Editing}

\subsubsection*{D1. Definition of Evaluation Metrics}

\textbf{Fidelity} refers to the accuracy with which the edited audio preserves the original content, specifically the degree to which the audio segments that do not require editing are retained. \textbf{Perceptual quality} refers to the evaluation of the overall auditory experience of the edited audio, with a focus on its naturalness, clarity, and lossless sound quality. \textbf{Consistency} refers to the evaluation of the smoothness of transitions between different processed segments, such as the shift between edited and unedited parts in audio. \textbf{Regional specificity} refers to whether the model only performs editing within the designated region without affecting other parts of the audio. \textbf{Instruction adherence} refers to evaluating whether the model operates in accordance with the specific editing instructions provided by the user. We selected 20 edited audio clips from each of the categories of addition, deletion, and replacement, conducted manual scoring on a 100-point scale, and took the average value as the subjective evaluation result.

\subsubsection*{D2. Visualization of Edited Sound Effect}

Figure 4 shows the visualization results of the sound effect editing performed by AV-Edit. The left part of Figure 4 shows the editing manipulation of addition. The input audio is a segment of ocean wave sounds, and the input visuals include ocean waves and seagulls in the air. The model generates seagull sounds based on the semantics of the audio and visual, and the edited audio is obtained by combining these with the ocean wave sounds. The middle part of Figure 4 shows the editing manipulation of removal. The input audio includes guitar sound and castanet sound, while the input visuals only retain the action of playing the guitar. Based on the semantics of the audio and visual, the model removes the castanets sounds and outputs the edited guitar sounds. The right part of Figure 4 shows the editing manipulation of replacement. The input audio is a cat meowing, and the input visual is a dog. Based on the semantics of the audio and visual, the model replaces the cat's meow with a dog's bark.

\subsection*{E. More ablation experiments}

\subsubsection*{E1. Replacement of audio and visual encoders}

Replacing the audio and video encoders with the currently mainstream encoders to verify the effectiveness of the proposed audio-video co-encoder CAV-MAE-Edit. For audio and visual encoders from different models, we use CLAP and CLIP to replace our audio and visual encoders; for audio and visual encoders from the same model, we use ImageBind. Tables 1 and 2 show the impact of different encoders on audio generation and sound effect editing performance, and it can be seen that our encoder has comprehensive advantages.

\subsubsection*{E2. Delation of audio-visual correlation calculation}

Tables 3 and 4 illustrate the impact of removing the step of correlation calculation on the performance of audio generation and sound effect editing. It can be clearly observed that the correlation calculation strategy is both necessary and effective.

\begin{figure*}[htbp]
    \renewcommand{\thefigure}{A}
	\centering
	\includegraphics[width=0.9\textwidth]{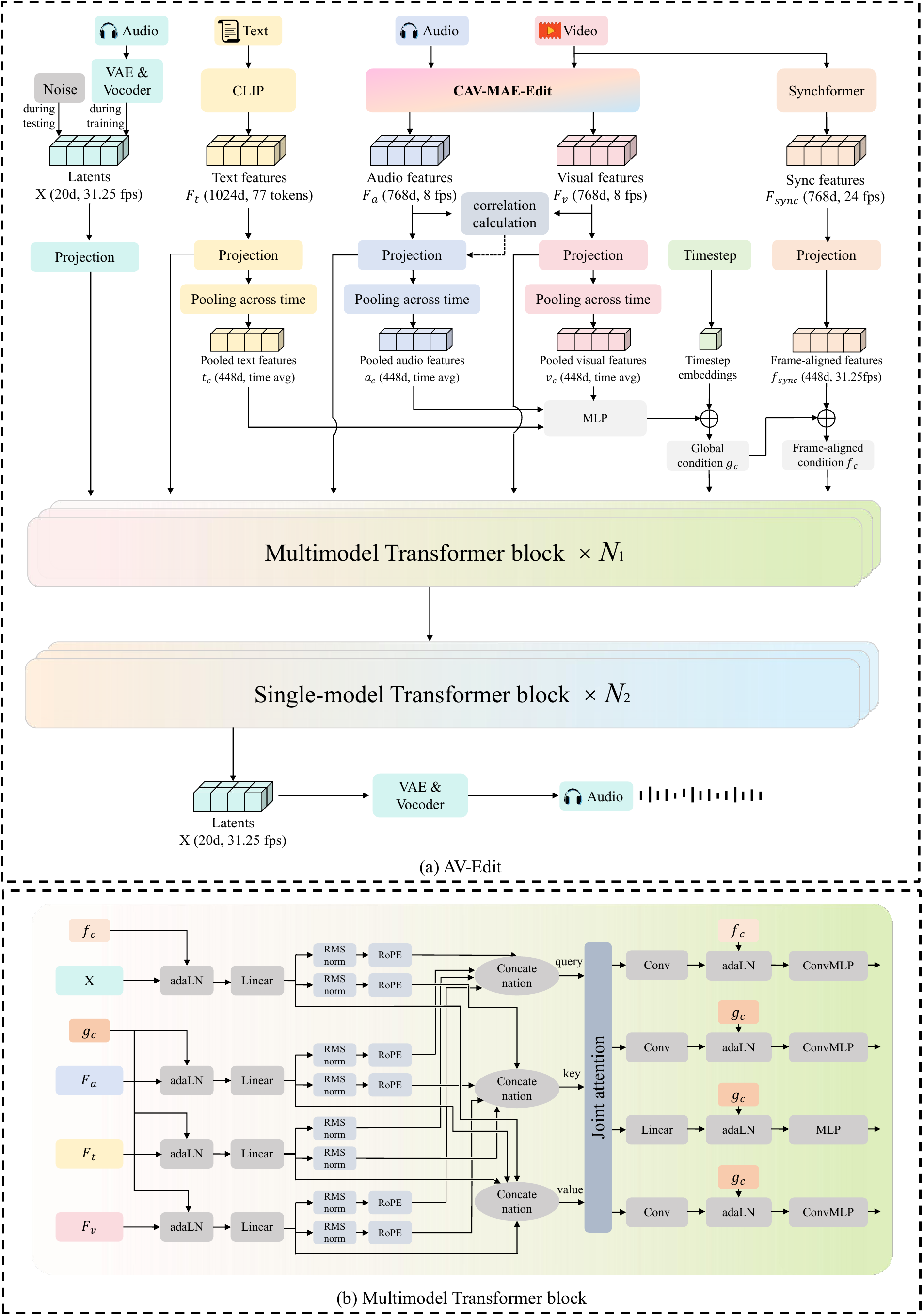} % Reduce the figure size so that it is slightly narrower than the column.
	\caption{Detailed structure of the generation section in AV-Edit.}
	\label{figA}
\end{figure*}

\begin{figure*}[htbp]
    \renewcommand{\thefigure}{C}
	\centering
	\includegraphics[width=1.0\textwidth]{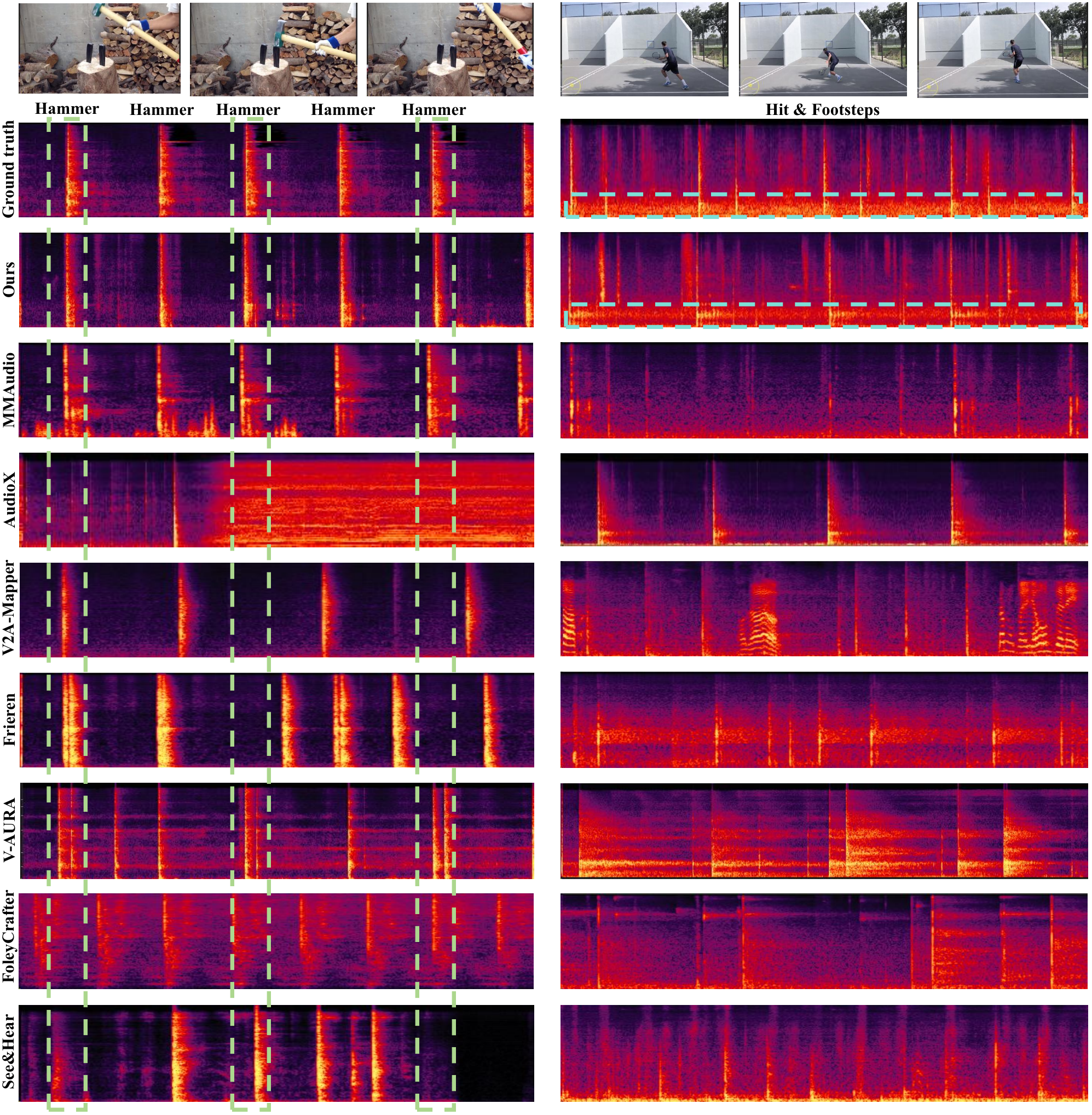} % Reduce the figure size so that it is slightly narrower than the column.
	\caption{The spectrograms of the ground truth and the generated audios.}
	\label{figC}
\end{figure*}

% \newpage
% \input{ReproducibilityChecklist}

\end{document}